\documentclass[journal]{IEEEtran}
\usepackage{amsmath}
\usepackage{graphicx}
\usepackage{amsfonts}
\usepackage{booktabs}
\usepackage{url}
\usepackage{cite}
\usepackage{amssymb}
\usepackage{diagbox}
\usepackage{subfigure}
\usepackage{comment,soul}
\usepackage{color}

\title{Mask-based Neural Beamforming for Moving Speakers with Self-Attention-based Tracking}
\author{Tsubasa Ochiai~\IEEEmembership{Member,~IEEE}, Marc Delcroix~\IEEEmembership{Senior Member,~IEEE},\\ Tomohiro Nakatani~\IEEEmembership{Fellow,~IEEE}, Shoko Araki~\IEEEmembership{Fellow,~IEEE}
\thanks{The authors are with NTT Corporation, Japan.}}

\begin{document}
%
\maketitle
\begin{abstract}

Beamforming is a powerful tool designed to enhance speech signals from the direction of a target source. Computing the beamforming filter requires estimating spatial covariance matrices (SCMs) of the source and noise signals. Time-frequency masks are often used to compute these SCMs. Most studies of mask-based beamforming have assumed that the sources do not move. However, sources often move in practice, which causes performance degradation. In this paper, we address the problem of mask-based beamforming for moving sources. We first review classical approaches to tracking a moving source, which perform online or blockwise computation of the SCMs. We show that these approaches can be interpreted as computing a sum of instantaneous SCMs weighted by attention weights.
These weights indicate which time frames of the signal to consider in the SCM computation. Online or blockwise computation assumes a heuristic and deterministic way of computing these attention weights that, although simple, may not result in optimal performance. We thus introduce a learning-based framework that computes optimal attention weights for beamforming. We achieve this using a neural network implemented with self-attention layers. We show experimentally that our proposed framework can greatly improve beamforming performance in moving source situations while maintaining high performance in non-moving situations, thus enabling the development of mask-based beamformers robust to source movements.

\end{abstract}

\IEEEpeerreviewmaketitle
\section{Introduction}
\label{sec:intro}

\IEEEPARstart{M}ICROPHONE array signal processing \cite{van1988beamforming,brandstein2001microphone,benesty2008microphone}, which uses spatio-temporal information obtained with multiple microphones, has been an active research field for several decades and plays an important role in the development of many applications. In particular, multichannel linear filtering using a microphone array, i.e., beamforming, has been used extensively to design speech enhancement systems for hearing aids \cite{doclo2010acoustic,doclo2007frequency} and for noise-robust automatic speech recognition (ASR) systems \cite{haeb2020far,boeddeker2018exploring,heymann2018performance}.
Recently, the mask-based beamforming approaches \cite{heymann2016neural,erdogan2016improved,higuchi2016robust} have attracted increased attention because they were shown to be particularly effective in reducing noise or the effect of interference speakers in recent robust ASR challenges \cite{ barker2015third, barker2018fifth}. 

A beamformer exploits the spatial information about the target and interfering sources derived from spatial covariance matrices (SCMs) to emphasize the signals coming from a target source direction while suppressing the interfering signals.
The mask-based beamformer exploits time-frequency masks derived from neural networks (NNs) \cite{heymann2016neural,erdogan2016improved} or other source models such as complex Gaussian mixture models (cGMMs) \cite{higuchi2016robust} to compute the SCMs. SCMs capture the spatial information and are thus sensitive to source movements. Most studies involving mask-based beamformers avoided this issue by assuming that the target and interfering sources do not move within an utterance. However, this hypothesis may not hold in general, especially when considering more realistic situations such as sound captured by a smart speaker or robots, where the target speaker or interference speakers could, for example, walk around the room while talking.
In this paper, we address the problem of designing a mask-based beamformer that is robust to moving sources by proposing a novel estimation framework of the beamforming filters that can track the source movements.

Mask-based beamformers compute the source and interference SCMs by averaging over time the outer product of the multi-channel observation vectors (i.e., the vector of the multi-channel observed signal at each time-frequency bin) masked with the time-frequency masks. We can compute \emph{time-invariant} SCMs over an entire utterance if we assume that the sources are not moving.
This procedure results in a time-invariant beamformer. 

Adapting this framework to a moving source scenario requires estimation of \emph{time-varying} SCMs and beamforming filters, which reflect changes in the acoustic conditions, i.e., the source positions. We can estimate the time-varying SCMs using online or blockwise processing.
For example, the online mask-based beamformer \cite{higuchi2016robust,boeddeker2018exploring,malek2019block,togami2019simultaneous} sequentially updates the SCMs.
These approaches estimate one SCM and the resultant beamforming filters for each frame or block, not for the entire utterance, and thus they could potentially deal with moving sources. However, they require tuning hyperparameters, such as the forgetting factor and block size that may vary depending on, e.g., the speed of the sources.
Consequently, these approaches may not track a source in an optimal way.

We can view the computing of SCMs by online or blockwise processing as limiting the range of frames that contribute to estimating the SCMs for each frame or block.
That is, such processing replaces the averaging operation in the SCM computation with a weighted averaging, where the weights indicate the range of time frames to consider for the computation of the SCMs.
We call these weights \emph{attention weights}.
Conventional online or blockwise processing use a simple heuristic rule to determine the attention weights.
We propose improving upon this heuristic rule by introducing a novel framework that automatically determines the optimal attention weights based on an NN.
Concretely, we design an NN that accepts the observed signals and the time-frequency masks as the inputs and predicts attention weights that determine which time frames to focus on when computing the SCMs at a given time.
This mechanism can be implemented naturally using the attention mechanism that has been widely used in many machine learning applications \cite{vaswani2017attention,dong2018speech,dosovitskiy2020image}.
We train the NN by minimizing a loss computed between the target source of a moving speaker and the output of a time-varying beamformer, which employs the time-varying SCMs computed with the attention weights.
With this fully supervised scheme, we can learn to predict optimal attention weights that allow the beamforming to steer its directivity toward the position of the moving source for each frame, i.e., that enable implicit source tracking by the attention mechanism.

Note that \emph{time-varying} beamformers are often investigated for online (sequential) systems that target processing with low latency, but they can also be used for offline systems to estimate better SCMs and beamforming filters such as those in a previous work \cite{kubo2019mask}.
Similarly, in this paper, we focus on offline processing that utilizes all of the information within an utterance.
We could easily extend the proposed framework to sequential processing by restricting the use of future frames, but this is out of the scope of this paper.

We tested the effectiveness of the proposed framework on moving source signals simulated using the Wall Street Journal (WSJ0) corpus \cite{paul1992design} for the speech signals, dynamic room impulse responses computed with the gpuRIR toolkit \cite{diaz2021gpurir}, and background noise derived from CHiME-3 corpus \cite{barker2015third}.
Experimental results show that the proposed framework achieves better speech enhancement and ASR performance, i.e., signal-to-distortion ratio (SDR), perceptual evaluation of speech quality (PESQ), short-time objective intelligibility (STOI), and word error rate (WER), compared to the conventional time-invariant, online, and blockwise beamforming frameworks.
In addition, we confirmed that our proposed scheme could track a moving source by visualizing the directivity characteristics (i.e., beam patterns) of the time-varying beamformer computed with our proposed scheme.

The main contributions of this paper are as follows:
\begin{enumerate}
    \item We propose a fully supervised scheme to allow the design of time-varying mask-based beamformers that can track moving sources.
    \item We introduce an self-attention-based NN that predicts the time frames that are relevant for computing the SCMs at a given time.
    \item We design an experiment using simulated moving sources to compare the different approaches for tackling moving sources and show the superiority of our proposed framework for both speech enhancement and ASR.
\end{enumerate}

The remainder of this paper is summarized as follows.
In Section~\ref{sec:prior}, we briefly discuss prior works related to our approach.
Section~\ref{sec:conventional} describes the conventional mask-based beamforming framework.
In Section~\ref{sec:proposed}, we first generalize the online and blockwise framework and then introduce the proposed time-varying beamforming framework with the attention weight estimation model.
In Section~\ref{sec:experiment}, we detail the experimental conditions of the moving source scenario and demonstrate the effectiveness of the proposed framework.
Finally, we conclude this paper in Section~\ref{sec:conclusion}.

\section{Related works}
\label{sec:prior}

Here, we briefly review related speech enhancement approaches that deal with source movements. 
\subsection{Beamformer-based approach}
\subsubsection{Mask-based beamformer}

A mask-based beamformer first computes a time-frequency mask, which indicates the time-frequency bins where the target source is dominant. The mask is used to compute the SCMs of the target source and noise, which are required to compute the beamformer coefficients.

There are currently two main research directions toward estimating the time-frequency masks for mask-based beamformers, i.e., spatial clustering \cite{higuchi2016robust} and NNs \cite{heymann2016neural,erdogan2016improved}.
The spatial clustering-based approaches estimate the time-frequency masks based on the spatial information, which is derived from the microphone array signals, and thus the estimation accuracy is affected by the movements of the source signals.
On the other hand, the NN-based approaches estimate the time-frequency masks mainly based on the spectral information, which can be derived even from a single microphone signal, and thus, in principle, these methods are not affected by source movements.
Therefore, we adopt the NN-based approach to estimate the time-frequency masks of moving sources.

Many related studies have investigated online/low-latency processing for mask-based beamformers, e.g., \cite{higuchi2016robust,boeddeker2018exploring,malek2019block,togami2019simultaneous}.
Most of these studies focused on the online computation of the beamformer coefficients given the masks.
However, only a few approaches have actually been evaluated with moving source scenarios.
For example, in a prior work \cite{malek2019block}, the authors introduced the block-online processing of a mask-based beamformer to deal with a moving source scenario.

Other work \cite{togami2019simultaneous} investigated using an NN to predict the forgetting factor for online computation of the SCM, but it was not evaluated on moving source scenarios.
Our approach can be considered the generalization of that previous effort \cite{togami2019simultaneous}, where we extend the formalization to offline processing and introduce self-attention-based NNs that naturally generalize the computation of the time-varying SCMs of conventional online and blockwise approaches.
Furthermore, we evaluated and analyzed the behavior of the proposed approach on a moving source dataset.

Time-varying mask-based beamformers have also been investigated to improve performance for offline processing.
For example, our previous effort assumed a time-varying noise covariance matrix in designing a time-varying beamformer that could adapt \emph{to variations in noise conditions} \cite{kubo2019mask}.
Nevertheless, although this beamformer is time-varying, it is not designed to handle moving sources.

\subsubsection{Other types of neural beamformers}

Besides the mask-based beamformer, there are currently two main research directions toward estimating the beamforming filters with NNs (i.e., neural beamformer).
One approach consists of directly estimating the time-varying beamforming filters as an NN's output, e.g., \cite{xiao2016deep,zhang2021adl,xu2021generalized}.
The other approach uses a set of fixed time-invariant beamforming filters implemented as a layer of an NN and an integration layer that combines the beamformer outputs for each time frame, e.g., \cite{sainath2016reducing,li2021deep}.
Such neural beamformers were integrated into ASR systems, and their parameters (i.e., the fixed beamformer and integration layer) were jointly optimized during the training of the ASR system. 

Both of these approaches achieve time-varying beamforming and thus have the potential to handle moving source scenarios.
However, they have been evaluated mostly for non-moving situations or for a sudden change of the source position within an utterance \cite{li2021deep}.
Moreover, the beamforming filters are black boxes that are optimized directly from multi-channel data, and thus, it is difficult to include physical knowledge (e.g., constraints) from microphone array signal processing theory.
Furthermore, they may be dependent on the microphone array geometry used during training, and the filter size (number of channels used for beamforming) is fixed by the number of channels at the input and output of the NN.

In contrast to such types of neural beamformers, mask-based beamformers combine the high estimation capability of NNs with physical knowledge from microphone array signal processing theory, such as the distortionless constraint of the Minimum Variance Distortionless Response (MVDR) beamformer.
Moreover, they do not rely on the microphone array geometries (such as number of channels, microphone locations, and ordering) because the NN used to predict the time-frequency mask can be trained on single-channel data.
Consequently, the mask-based beamformer with distortionless constraint has become a de facto standard to construct robust ASR systems in recent noisy ASR challenges (e.g., CHiME-3 to CHiME-6\cite{barker2015third,barker2018fifth}).
Motivated by the successes and advantages of the mask-based beamformers, in this study, we focus on the extension of the mask-based beamformer scheme to handle moving sources.

\subsection{Source localization-based approach}

It is also possible to deal with moving sources by combining source localization/tracking with a beamformer designed to steer in the direction of the source \cite{chang2019adaptive}.
Source localization for the moving sources has been an active research field for several decades \cite{schmidt1986multiple,dibiase2001robust,chakrabarty2019multi,schymura2021pilot}.
Although great progress has been made in recent years, the accuracy of the source localization still tends to degrade in high reverberation and noise conditions \cite{evers2020locata}.
Moreover, it is challenging to track the sources when they move in silence.
These limitations may make precise localization challenging, which as a result would impact the performance of the beamformer.

Our proposed framework does not explicitly conduct source localization, which avoids the impact of localization errors on beamforming performance.

\subsection{Blind source separation approach}

Source movements are also a problem for microphone-array-based blind source separation.
Recently, several studies proposed estimating time-invariant separation filters that are robust to source movements \cite{jansky2022auxiliary,amor2021blind,koldovsky2021dynamic}. 
However, these approaches may deal with only relatively small source movements because the filters are time-invariant.

In contrast, our proposed framework estimates the frame-by-frame time-varying filters that can track a source even for large movements such as 360\textdegree movements in the experiments of Section~\ref{sec:beampattern}.

\section{Conventional mask-based beamformer}
\label{sec:conventional}

\subsection{Problem definition}

Let $\mathbf{Y}_{t,f} = \{Y_{t,f,c=1}, \ldots, Y_{t,f,c=C} \} \in \mathbb{C}^{C}$ be a vector comprising the $C$-channel short-time Fourier transform (STFT) coefficients of the observed noisy signal at a time-frequency bin $(t,f)$, where $Y_{t,f,c} \in \mathbb{C}$ is the STFT coefficient for the $c$-th channel. Let $T$ and $F$ be the number of time frames and frequency bins, respectively.
Assuming that the acoustic condition (i.e., the transfer function) is static within a short-time duration (i.e., a short time frame), the observed signal $\mathbf{Y}_{t, f} \in \mathbb{C}^{C}$ can be approximately modeled as:
\begin{align}
    \mathbf{Y}_{t, f} = \mathbf{H}_{t,f} S_{t,f} + \mathbf{N}_{t, f},
    \label{eq:obs_model}
\end{align}
where $S_{t,f} \in \mathbb{C}$ and $\mathbf{N}_{t, f} \in \mathbb{C}^{C}$ denote the speech source and additive noise signals at the time-frequency bin $(t, f)$, respectively.
$\mathbf{H}_{t,f} \in \mathbb{C}^{C}$ denotes the time-varying transfer function between the speech source and the microphones at a time-frequency bin $(t, f)$.

When the source is not moving, we can assume that the transfer function is static within an utterance, i.e., $\mathbf{H}_{t,f} = \mathbf{H}_{f}$, and thus use time-invariant beamformers to enhance the noisy speech signals.
This is the scheme used in many studies and challenges \cite{barker2015third,barker2018fifth}.
However, in general, the transfer function dynamically changes due to, e.g., the movements of the source, which is the situation we tackle in this paper. Therefore, we assume the observation model of Eq.~\eqref{eq:obs_model} and investigate the design of time-varying (frame-by-frame) mask-based beamformers to enhance the speech source $S_{t,f}$.

\subsection{Minimum variance distortionless response beamformer}
\label{sec:mvdr}

Given the observed noisy signal $\mathbf{Y}_{t,f}$, a frequency-domain beamformer estimates the STFT coefficient of the enhanced speech, $\hat{S}_{t,f} \in \mathbb{C}$, as follows:
\begin{align}
    \hat{S}_{t,f} = \mathbf{w}_{t,f}^{\mathsf{H}} \mathbf{Y}_{t,f}, \label{eq:bf}
\end{align}
where $\mathbf{w}_{t,f} \in \mathbb{C}^{C}$ denotes a vector comprising the beamforming filter coefficients and $\empty^{\mathsf{H}}$ represents the conjugate transpose.
We then obtain the time-domain enhanced signal, $\hat{\mathbf{s}} \in \mathbb{R}^{\mathcal{T}}$, by applying the inverse STFT to $\hat{S}_{t,f}$ and the overlapping add method, where $\mathcal{T}$ denotes the duration of the time-domain signal.

We can compute the beamforming filter coefficients from the SCMs of the speech and noise signals.
We adopt in this paper a widely used MVDR formalization, which computes the beamforming filter coefficients $\mathbf{w}_{t, f}$ as follows \cite{souden2009optimal}:
\begin{align}
    \mathbf{w}_{t, f} = \frac{(\mathbf{\Phi}^{\text{N}}_{t, f})^{-1} \mathbf{\Phi}^{\text{S}}_{t, f}}{\text{Tr}((\mathbf{\Phi}^{\text{N}}_{t, f})^{-1} \mathbf{\Phi}^{\text{S}}_{t, f})} \mathbf{u}, \label{eq:mvdr}
\end{align}
where $\mathbf{\Phi}^{\mathrm{S}}_{t, f} \in \mathbb{C}^{C \times C}$ and $\mathbf{\Phi}^{\mathrm{N}}_{t, f} \in \mathbb{C}^{C \times C}$ are the SCMs of the speech and noise signals at time-frequency bin $(t, f)$, respectively.
$\mathbf{u} \in \mathbb{R}^{C}$ is a one-hot vector representing the index of the reference microphone.

\subsection{Mask-based spatial covariance matrix estimation}
\label{sec:SCM_computation}

The mask-based beamforming scheme relies on the sparseness property of speech signals in the STFT domain \cite{wang2008time} to estimate the SCMs using time-frequency masks \cite{vu2010blind,souden2013multichannel,heymann2016neural,erdogan2016improved,higuchi2016robust}.
Here, the masks indicate the time-frequency bins where the source or noise is dominant.
In the following, we briefly overview several commonly used options for estimating the SCMs from the time-frequency masks.

\subsubsection{Time-invariant SCM computation}

Assuming that the transfer function is static within the utterance, we can compute the time-invariant SCMs $\mathbf{\Phi}^{\nu}_{f}$ as \cite{heymann2016neural,erdogan2016improved,higuchi2016robust}:
\begin{align}
    \mathbf{\Phi}^{\nu}_{f} =& \sum_{\tau=1}^{T} \frac{1}{\sum_{\tau^{'}=1}^{T} m^{\nu}_{\tau^{'},f}} \underbrace{m^{\nu}_{\tau,f} \mathbf{Y}_{\tau,f} \mathbf{Y}^{\mathsf{T}}_{\tau,f}}_{ \triangleq\mathbf{\Psi}^{\nu}_{\tau,f} }, \label{eq:ti_scm}
\end{align}
where $m_{t,f}^{\nu} \in [0, 1]$ is a time-frequency mask and $\nu \in \{\text{S}, \text{N}\}$ are the indexes for speech and noise, respectively. By abuse of terminology, we call $\mathbf{\Psi}^{\nu}_{t,f}$ the \emph{instantaneous} SCM (ISCM) at time-frequency bin $(t,f)$.
Because the SCMs are time-invariant, the beamforming filter coefficients computed with Eq.~\eqref{eq:mvdr} are also time-invariant.
Therefore, this approach cannot handle moving sources well.

\subsubsection{Online SCM computation}
\label{sec:onl}

A conventional way to compute a time-varying SCM $\mathbf{\Phi}^{\nu}_{t,f}$ is to use a recursive approach \cite{higuchi2016robust,boeddeker2018exploring,malek2019block,togami2019simultaneous}:
\begin{align}
    \mathbf{\Phi}^{\nu}_{t,f} = & \alpha \mathbf{\Phi}^{\nu}_{t-1,f} + \mathbf{\Psi}^{\nu}_{t,f} \\
     = & \sum_{\tau=1}^{t} \alpha^{t-\tau} \mathbf{\Psi}^{\nu}_{\tau,f}, \label{eq:onl} 
\end{align}
where $\alpha$ denotes the forgetting factor, which gives exponentially less weight to the older ISCMs. With this approach, the SCMs and the beamforming filter coefficients are estimated at each time frame, which would allow tracking a source. However, the tracking speed depends on the forgetting factor.
It may thus be challenging to tune this parameter to offer optimal performance for various conditions of source movement.

In this paper, we adopt the frame-by-frame update of the beamforming filters for the online processing to allow precise tracking instantaneously.

\subsubsection{Blockwise SCM computation}

An alternative way of computing time-varying SCMs is to use blockwise processing \cite{kubo2019mask}, i.e., dividing a signal into consecutive time blocks and computing the SCMs for each block as follows:
\begin{align}
    \mathbf{\Phi}^{\nu}_{t,f} = \sum_{\tau=t-L}^{t+L} \frac{1}{\sum_{\tau^{'}=t-L}^{t+L} m^{\nu}_{\tau^{'},f}} \mathbf{\Psi}^{\nu}_{\tau,f}, \label{eq:blk} 
\end{align}
where $L$ is a block size parameter that denotes the half span of the blocks, and thus $2L + 1$ frames are used for the SCM computation of each block.

Setting the block size requires a trade-off between using a large block size to allow computing reliable statistics and a small block size to allow better tracking.
Therefore, as with the online SCM computation, tuning this parameter may be challenging and lead to sub-optimal performance.

In addition to the block size, we can consider the block shift, which determines how often we compute the SCMs and beamforming filter coefficients.
In the experiments of Section~\ref{sec:experiment}, we use a block shift of one frame, which means that the SCMs and the beamforming filter coefficients are computed for each frame like the online SCM computation described in Section~\ref{sec:onl}.

\section{Proposed time-varying SCM computation with self-attention-based weighting}
\label{sec:proposed}

\subsection{Generalized formulation of SCM computation}

We can express the different SCM computation approaches using a general formulation as:
\begin{align}
    \mathbf{\Phi}^{\nu}_{t,f} = \sum_{t^{'}=1}^{T} c_{t,t^{'}}^{\nu} \mathbf{\Psi}^{\nu}_{t^{'},f}, \label{eq:scm}
\end{align}
where $\mathbf{c}^{\nu}_{t} = \{ c^{\nu}_{t,t^{'}=1}, \ldots, c^{\nu}_{t,t^{'}=T}\} \in \mathbb{R}^{T}$ are weight coefficients that control the range for accumulating the statistics used to compute the SCMs for the $t$-th frame. 
The weight coefficients $\mathbf{c}^{\nu}_{t}$ determine which time frames to focus on when computing the SCMs at a given time frame (i.e., $t$) among all time frames (i.e., $t^{'}=1 \sim T$).
In this paper, we refer to these weight coefficients as attention weights.

We can easily see that the SCM computation approaches discussed in Section~\ref{sec:SCM_computation} are special cases of the general formulation of Eq.~\eqref{eq:scm}. The time-invariant computation of Eq.~\eqref{eq:ti_scm} corresponds to, 
\begin{align}
    c^{\nu}_{t,t^{'}} = \frac{1}{\sum_{\tau^{'}=1}^{T} m^{\nu}_{\tau^{'},f}},
\end{align}
the online computation of Eq~\eqref{eq:onl} to,
\begin{align}
    c^{\nu}_{t,t^{'}} = 
    \begin{cases}
        \alpha^{t-t^{'}} & t^{'} \le t, \\
        0 &t^{'} > t,
    \end{cases}
\end{align}
and the blockwise computation of Eq.~\eqref{eq:blk} to setting,
\begin{align}
    c^{\nu}_{t,t^{'}} = 
    \begin{cases}
        \frac{1}{\sum_{\tau^{'}=t-L}^{t+L} m^{\nu}_{\tau^{'},f}} & t^{'} \in [t-L, \ldots, t+L], \\
        0 &t^{'} \notin [t-L, \ldots, t+L].
    \end{cases}
\end{align}

\subsection{Self-attention-based time-varying attention weight estimation}
\label{sec:proposed_proc}

\subsubsection{Overall procedure of time-varying attention weight estimation}

As mentioned above, the online and blockwise SCM computations use simple rules to compute the attention weights.
These approaches would allow handling moving source scenarios, but such simple rules may not be necessarily optimal for tracking moving sources.
In this section, we propose instead to design an NN to estimate optimal attention weights.

\begin{figure}[t]
  \centering
  \includegraphics[width=0.8\linewidth]{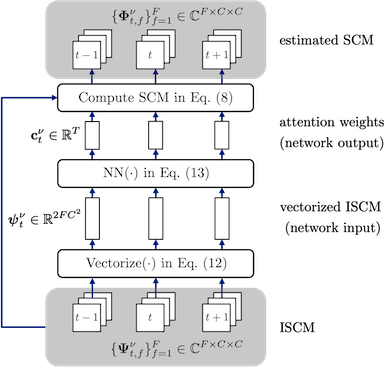}
  \caption{Overview of estimation procedure of time-varying SCMs with attention weight estimation neural network.}
  \label{fig:proposed}
\end{figure}

Figure~\ref{fig:proposed} illustrates the proposed estimation procedure of the time-varying SCMs with self-attention-based weighting.
The method relies on an NN that accepts the ISCMs' coefficients for the entire signal and predicts optimal attention weights. We detail the process below.

First, to make the ISCMs suitable for the NN's input, we convert the ISCMs of all frequency bins at a given time frame $t$ $\{ \boldsymbol{\Psi}^{\nu}_{t,f} \}_{f=1}^{F} \in \mathbb{C}^{F \times C \times C}$ into a real-valued vector $\boldsymbol{\psi}^{\nu}_{t} \in \mathbb{R}^{2FC^{2}}$ as:
\begin{align}
    \boldsymbol{\psi}^{\nu}_{t} = \text{Vectorize}(\{\mathbf{\Psi}^{\nu}_{t,f}\}_{f=1}^{F}),
\end{align}
where $\text{Vectorize}(\cdot)$ represents the unfolding operation that converts the complex-valued tensor $\{ \boldsymbol{\Psi}^{\nu}_{t,f} \}_{f=1}^{F}$ into the real-valued vector $\boldsymbol{\psi}^{\nu}_{t}$, which contains the real and imaginary parts of all elements of the tensor.

We use the sequence of vectorized ISCMs $\{ \boldsymbol{\psi}^{\nu}_{t} \}_{t=1}^{T}$ as input to an NN that estimates the time-varying attention weight coefficients $\{\mathbf{c}_{t}^{\nu}\}_{t=1}^{T}$ as follows:
\begin{align}
    \{\mathbf{c}_{t}^{\nu}\}_{t=1}^{T} = \text{NN}^{\nu}(\{\boldsymbol{\psi}^{\nu}_{t}\}_{t=1}^{T}; \mathbf{\Lambda^{\nu}}), \label{eq:nn}
\end{align}
where $\text{NN}^{\nu}(\cdot)$ is the non-linear transformation of an NN and $\mathbf{\Lambda^{\nu}}$ denotes the learnable parameters of $\text{NN}^{\nu}(\cdot)$.
$\text{NN}^{\nu}(\cdot)$ should predict attention weights that allow accumulating ISCMs from a similar direction to estimate reliable SCMs while making it possible to track a moving source.
Since the input ISCMs capture information about the source direction, this behavior can naturally be implemented using an architecture for $\text{NN}^{\nu}(\cdot)$ inspired by self-attention network \cite{vaswani2017attention}, which estimates the weight coefficients focusing on the similarity between the input frames.

\begin{figure}[t]
  \centering
  \includegraphics[width=0.8\linewidth]{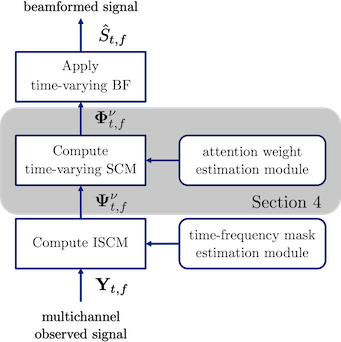}
  \caption{Overall procedure of proposed time-varying beamforming system, which is constructed from the time-varying beamformer with the time-frequency mask and attention weight estimation modules.}
  \label{fig:overview}
\end{figure}

Figure~\ref{fig:overview} summarizes the overall procedure of our proposed time-varying beamforming system, which consists of the time-frequency mask and attention weight estimation modules.
First, with the mask estimation module, we estimate the time-frequency masks $m^{\nu}_{t,f}$ and compute the ISCMs $\mathbf{\Psi}^{\nu}_{t,f}$ defined in Eq.~\eqref{eq:ti_scm}.
Then, with the attention weight estimation module, we estimate the attention weights $c_{t,t^{'}}^{\nu}$ and compute the time-varying SCMs with Eq.~\eqref{eq:scm}.
Finally, we construct the time-varying beamformer based on Eq.~\eqref{eq:mvdr} and obtain the enhanced signals $\hat{S}_{t,f}$ with Eq.~\eqref{eq:bf}.

\subsubsection{Overview of attention module}
\label{sec:proposed_att}

Here, we briefly review the formulation of an attention module.
Let $\mathbf{q}_{t} \in \mathbb{R}^{1 \times D^{\mathrm{KQ}}}$, $\mathbf{k}_{t} \in \mathbb{R}^{1 \times D^{\mathrm{KQ}}}$, and $\mathbf{v}_{t} \in \mathbb{R}^{T \times D^{\mathrm{V}}}$ be the vectors at time frame $t$ called query, key, and value, respectively.
Here, $D^{\mathrm{KQ}}$ denotes the dimension of the query and key, and $D^{\mathrm{V}}$ denotes the dimension of the value.

Given the sequence of the queries $\mathbf{Q} = \{ \mathbf{q}_{t=1}, \ldots, \mathbf{q}_{t=T} \} \in \mathbb{R}^{T \times D^{\mathrm{KQ}}}$, keys $\mathbf{K} = \{ \mathbf{k}_{t=1}, \ldots, \mathbf{k}_{t=T} \} \in \mathbb{R}^{T \times D^{\mathrm{KQ}}}$, and values $\mathbf{V} = \{ \mathbf{v}_{t=1}, \ldots, \mathbf{v}_{t=T} \} \in \mathbb{R}^{T \times D^{\mathrm{V}}}$ as a matrix form, the output of the attention module is computed as:
\begin{align}
\mathbf{A} &= \text{Weight}(\mathbf{Q}, \mathbf{K})
           = \text{Softmax} \Bigl(\frac{\mathbf{Q} \mathbf{K}^{\mathsf{T}}}{\sqrt{D^{\mathrm{KQ}}}} \Bigr), \label{eq:att_weight} \\
           \mathbf{Z} &= \text{Att}(\mathbf{A}, \mathbf{V})
           = \mathbf{A} \mathbf{V}, \label{eq:att}
\end{align}
where $\text{Weight}(\mathbf{Q}, \mathbf{K})$ denotes the function computing the attention weights $\mathbf{A} = \{ \mathbf{a}_{t=1}, \cdots, \mathbf{a}_{t=T} \} \in \mathbb{R}^{T \times T}$, and $\text{Att}(\mathbf{A}, \mathbf{V})$ denotes the function computing the attention output $\mathbf{Z} = \{ \mathbf{z}_{t=1}, \cdots, \mathbf{z}_{t=T} \} \in \mathbb{R}^{T \times D^{\mathrm{V}}}$.
Here, $\mathbf{a}_{t} = \{ a_{t, t^{'}=1} \ldots, a_{t, t^{'}=T} \} \in \mathbb{R}^{1 \times T}$ and $\mathbf{z}_{t} \in \mathbb{R}^{1 \times D^{\mathrm{V}}}$ denote the attention weight and output corresponding to query time $t$, respectively.
$\text{Softmax}(\cdot)$ is the softmax function \cite{goodfellow2016deep} that normalizes the attention weights over a key's axis.
A \emph{self-attention} module is a special case of attention that uses the same features for the query, key, and values.

The attention module outputs the sum of the value features weighted by the attention weights as in Eq.~\eqref{eq:att}.
We can confirm that the computation of Eq.~\eqref{eq:att} is similar to that of Eq.~\eqref{eq:scm} because we can reformulate Eq.~\eqref{eq:att} in a vector form as $\mathbf{z}_{t} = \sum_{t^{'}=1}^{T} a_{t, t^{'}} \mathbf{v}_{t^{'}}$, where the value $\mathbf{v}_{t^{'}}$ in Eq.~\eqref{eq:att} corresponds to the ISCM $\mathbf{\Psi}^{\nu}_{t^{'},f}$ in Eq.~\eqref{eq:scm}, the attention weight $a_{t, t^{'}}$ corresponds to the weight $c_{t,t^{'}}^{\nu}$, and the output $\mathbf{z}_{t}$ corresponds to the estimated SCM $\mathbf{\Phi}^{\nu}_{t,f}$.

Moreover, as seen from Eq.~\eqref{eq:att_weight}, the attention module determines the attention weights $a_{t, t^{'}}$ based on the dot-product similarity between queries and keys, and thus, the weight values become large when the input query and key features are similar.
Therefore, the attention module would give larger weight values to the time frames where the positions of the target source speaker are similar, and it could thus automatically determine the frame regions suitable for computing the time-varying SCMs $\mathbf{\Phi}^{\nu}_{t,f}$ considering the position of the moving source speakers. Consequently, the self-attention-based NN can perform source tracking implicitly.

In more detail, to increase the representation capability, we adopted a stacked self-attention architecture \cite{vaswani2017attention}, which consists of multiple self-attention modules as follows:
\begin{align}
    \mathbf{Z}_0 &= \{\boldsymbol{\psi}^{\nu}_{t}\}_{t=1}^{T}, \\
    \mathbf{A}_i &= \text{Weight}(\mathbf{K}{=} \mathbf{Z}_{i-1} \mathbf{W}^K_i, \mathbf{Q}{=} \mathbf{Z}_{i-1} \mathbf{W}^Q_i), \\
    \mathbf{Z}_i &= \begin{cases}
                    \text{Att}(\mathbf{A}_i, \mathbf{V}{=} \mathbf{Z}_{i-1} \mathbf{W}^V_i) & (i \neq I)\\
                    \text{Att}(\mathbf{A}_I, \mathbf{V}{=} \mathbf{Z}_{0}), & (i = I)
                    \end{cases}
\end{align}
where $\mathbf{Z}_0$ is the input representation of the NN, and $\mathbf{Z}_I$ is the output representation corresponding to the estimated time-varying SCMs $\mathbf{\Phi}^{\nu}_{t,f}$.
$\mathbf{Z}_i$ is the hidden representation at the $i$-th layer, and $I$ is the total number of layers.
Here, $\mathbf{W}^Q_i$, $\mathbf{W}^K_i$, and $\mathbf{W}^V_i$ are linear transformations associated with the query, key, and value, respectively.
The learnable parameters of $\text{NN}(\cdot)$ are $\mathbf{\Lambda} = \{ \mathbf{W}^Q_i, \mathbf{W}^K_i, \mathbf{W}^V_i \}_{i=1}^I$\footnote{To simplify the description, we explain the single-head attention case, although we use a multi-head attention followed by a position-wise feed-forward network \cite{vaswani2017attention} for $i \neq I$ in our experiments.}.

\subsection{Training procedure}

We train the attention weight estimation NN in an end-to-end manner with a mask-based beamformer so that it is possible to compute attention weights that are optimal for the beamforming of the moving source speaker; otherwise, it would be difficult to define the optimal target for the attention weights.
We assume that a set of input and target signals $\{\mathbf{y}, \mathbf{s}\}$ is available for training the model, where $\mathbf{y} \in \mathbb{R}^{\mathcal{T}}$ is the $\mathcal{T}$-length time-domain waveform of the observed noisy signal, and $\mathbf{s} \in \mathbb{R}^{\mathcal{T}}$ is its corresponding clean reverberant source signal.
As the training objective, we adopted the scale-dependent signal-to-noise ratio (SNR) \cite{roux2019sdr}.
The SNR loss $\mathcal{L}$ is expressed as follows:
\begin{align}
\mathcal{L} = - 10 \log_{10} \biggl( \frac{\| \mathbf{s} \|^{2}}{\| \mathbf{s} - \hat{\mathbf{s}} \|^{2}} \biggr), \label{eq:snr}
\end{align}
where $\hat{\mathbf{s}}$ denotes the time-domain waveform of the beamformed signal, which is computed based on the proposed scheme as described in Section~\ref{sec:proposed_proc}.

Through the training procedure, it is expected that the attention weight estimation networks learn to control the range for accumulating the ISCMs at each time step; consequently, the constructed beamformers can track the positions of the moving source speaker.
We incorporate various moving source conditions in the training set to learn robust tracking capabilities.
Such tracking behavior of the proposed scheme is visually analyzed in Section~\ref{sec:beampattern}.

\subsection{Weight Smoothing}
\label{sec:smooth}

Our preliminary experiments showed that while the proposed scheme is effective for improving the speech enhancement performance, e.g., SDR \cite{vincent2006performance}, it does not necessarily contribute to improving ASR performance.
We hypothesized that this is probably due to the non-smoothness introduced by the frame-by-frame processing.

To mitigate this issue, we introduce a scheme to smooth the attention weights estimated with the NN as:
\begin{align}
    \overline{\mathbf{c}}^{\nu}_{t} = \sum_{\tau=t-L^{'}}^{t+L^{'}} \frac{1}{2L^{'}+1} \mathbf{c}^{\nu}_{\tau}, \label{eq:smooth} 
\end{align}
where $\overline{\mathbf{c}}^{\nu}_{t}$ is the smoothed version of the weight coefficients and $L^{'}$ determines the number of frames used for the weight smoothing.

Here, Eq.~\eqref{eq:smooth} may look similar to the blockwise computation approach, since the summation of weights is performed over a window.
However, the weights $\mathbf{c}^{\nu}_{\tau}$ span the entire signal, unlike in blockwise processing, and it thus results in very different processing.

\section{EXPERIMENT}
\label{sec:experiment}

\subsection{Experimental conditions}

\begin{figure}[t]
  \centering
  \includegraphics[width=0.6\linewidth]{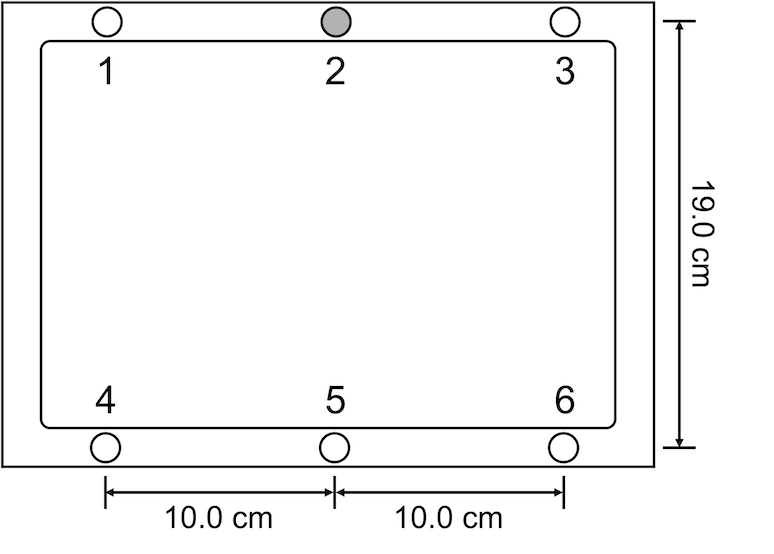}
  \caption{Microphone array geometry for CHiME-3 corpus. All microphones face forward except for microphone 2.}
  \label{fig:tablet}
\end{figure}

To evaluate the effectiveness of the proposed method, we created a new dataset of simulated moving sources in noisy conditions.
The signals for the speech source were taken from the WSJ0 corpus \cite{paul1992design} and those for the noise from the CHiME-3 corpus \cite{barker2015third}.
The CHiME-3 corpus contains noise signals recorded using a tablet device equipped with a rectangular microphone array with 6 channels, as illustrated in Figure~\ref{fig:tablet}.
From the 6-channel microphones, we excluded the second channel signals, which were captured by a microphone facing backward the tablet, and used the remaining five channels for the following multichannel experiments (i.e., $C=5$).

\begin{figure}[t]
  \centering
  \includegraphics[width=0.7\linewidth]{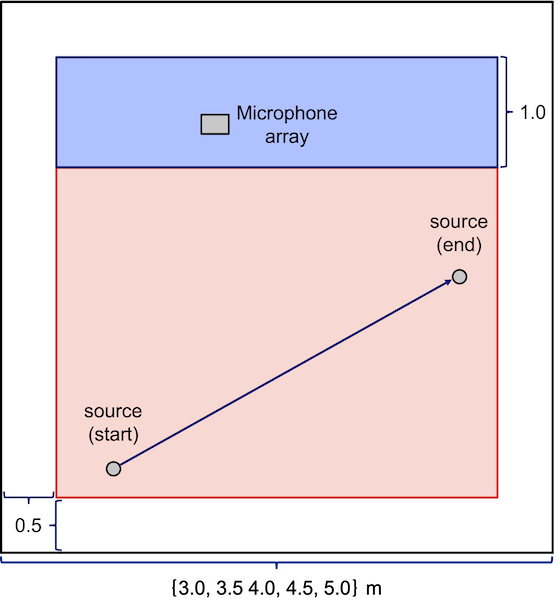}
  \caption{Simulation layout example of microphone array and moving source (moving in straight line)}
  \label{fig:simulation1}
\end{figure}

We randomly selected the pair of speech and noise signals from the WSJ0 and CHiME-3 corpora, respectively, and mixed them at various SNR between 2 dB and 8 dB.
We generated room impulse response (RIR) for moving sources using the gpuRIR simulation toolkit \cite{diaz2021gpurir}, which is based on the image method \cite{allen1979image}.
We used a randomly generated configuration (i.e., room geometry, array position, and source trajectory) for each simulated RIR. Figure~\ref{fig:simulation1} shows an example of such a layout.
In this experiment, we assumed that the room geometry was square and the source speaker was moving in a straight line in the room.
As illustrated in Figure~\ref{fig:simulation1}, the start and end positions of the source trajectory are randomly sampled from the red area, and the array position is randomly sampled from the blue area.
We set our simulation so that each moving speaker would start speaking an utterance at the start position and stop speaking at the end position.
The speed of the moving source speaker is constant within an utterance, but varies across utterances.
The reverberation time (T60) ranges from 0.1 to 0.3 s.
Table~\ref{tab:config1} summarizes the configuration of the moving source simulation.

We created 30,000, 2,000, and 2,000 noisy speech signals for training, development, and evaluation sets, respectively.
The speech sources for the training set were selected from WSJ0's training set ``si\_tr\_s.''
Those for the development and evaluation sets were selected from WSJ0's development set ``si\_dt\_05'' and evaluation set ``si\_et\_05,'' respectively.
We generate noisy signals using the noise from the CHiME-3 corpus.
We divided the noise sources in the CHiME-3 corpus into three subsets for training, development, and evaluation, containing 80 \%, 10\%, and 10\%, respectively, of the noise data of each environment (on a bus, in a cafe, pedestrian area, and street junction).

In addition to the above moving source dataset, we also created a non-moving source dataset as the additional evaluation set, which has exactly the same configuration as the moving source dataset (i.e., the pair of speech and noise sources and the RIR configurations) except that the source speaker position is fixed to the start position.

As the evaluation metrics, we used three speech enhancement measures; 1) the signal-to-distortion ratio (SDR) that permits time-invariant filters allowed distortions \cite{vincent2006performance}, 2) perceptual evaluation of speech quality (PESQ) \cite{rix2001perceptual}, and 3) short-time objective intelligibility (STOI) \cite{taal2011algorithm}; in addition, we used one speech recognition measure, i.e., word error rate (WER).
To compute the speech enhancement measures, we used the clean reverberant signals of the moving source speakers at the fifth channel as their references.

To evaluate the ASR performance, we created a deep neural network-hidden Markov model (DNN-HMM) hybrid ASR system \cite{hinton2012deep} based on Kaldi’s CHiME-4 recipe \cite{povey2011kaldi}.
The system was trained using the lattice-free maximum mutual information (MMI) criterion \cite{povey2016purely} with the noisy speech signals in the training set, and decoded with a trigram language model.
The details of the system are shown in Kaldi's recipe\footnote{\url{https://github.com/kaldi-asr/kaldi/tree/master/egs/chime4}}.

\begin{table}[t]
  \renewcommand{\arraystretch}{1.0}
  \caption{Configuration of moving source simulated data.}
  \label{tab:config1}
  \centering
  \scalebox{1.0}{
  \begin{tabular}{ l r }
    \toprule
    Corpus for source signal & WSJ0\\
    Corpus for noise signal & CHiME-3\\
    Number of microphones & 5\\
    Number of sources & 1\\
    Room width and depth & 3.0, 3.5, 4.0, 4.5, 5.0 m\\
    Room height & 2.5 m\\
    Reverberation time (T60) & 0.1 $\sim$ 0.3 s\\
    Source start/end positions & Random within red area of Figure~\ref{fig:simulation1}\\
    Source trajectory & Straight line \\
    Number of trajectory points & 32\\
    Signal-to-Noise Ratio & 2 $\sim$ 8 dB\\
    Height of microphones & 1.0 m\\
    Height of source & 1.5 $\sim$ 1.9 m\\
    \bottomrule
  \end{tabular}
  }
\vspace{-2mm}
\end{table}

\subsection{Experimental configurations}
\label{sec:expconf}

For the time-varying attention weight estimation module described in Section~\ref{sec:proposed}, we adopted a self-attention-based network architecture that is similar to the one used by the Transformer encoder \cite{vaswani2017attention}.
It consisted of stacked self-attention blocks, each of which was composed of the multi-head attention module followed by the position-wise feed-forward network.
For the training loss of the attention weight estimation NN, we adopted the SNR loss shown in Eq.~\eqref{eq:snr}, where the enhanced signals are obtained by applying the beamforming filters to the observed signals in Eq.~\eqref{eq:bf}.
The beamforming filter coefficients were obtained in Eq.~\eqref{eq:mvdr} using the time-varying SCMs estimated by the proposed method.
In the training stage, we used the ``Wiener like'' oracle time-frequency masks \cite{erdogan2015phase} to compute the ISCMs and optimized only the parameters of the attention weight estimation module ($\mathbf{\Lambda^{\nu}}$ in Section~\ref{sec:proposed_att}) based on the moving source dataset.
In the testing stage, we used the estimated time-frequency masks, which is obtained by averaging the estimated time-frequency masks computed from each microphone signal separately \cite{erdogan2016improved}.

For the time-frequency mask estimation module, we adopted a CNN-based network architecture \cite{bahmaninezhad2019comprehensive} that is similar to the time-domain audio separation network (TasNet) \cite{luo2019conv}.
It accepts a single-channel signal and outputs the time-frequency masks for the speech source.
The NN consists of stacked dilated convolution blocks. 
Unlike a previous related work \cite{luo2019conv}, it operates in the STFT domain \cite{bahmaninezhad2019comprehensive}.
For the training loss of the mask estimation NN, we adopted the SNR loss shown in Eq.~\eqref{eq:snr}, where the enhanced signals are obtained by applying the estimated time-frequency masks to the observed signals.

For the STFT computation, we used a Hanning window with a length and shift set at 64 ms and 16 ms, respectively.
The configurations related to the STFT and network architecture are briefly summarized in Table~\ref{tab:config2}, where we follow the notations introduced in \cite{vaswani2017attention} for attention weight estimation network and \cite{luo2019conv} for mask estimation network, respectively.

\begin{table}[t]
  \renewcommand{\arraystretch}{1.0}
  \caption{Summary of experimental configurations}
  \label{tab:config2}
  \centering
  \scalebox{1.0}{
  \begin{tabular}{ l r }
    \toprule
    Configuration of attention weight estimation network\\
    \midrule
    Number of attention heads ($d^{\text{head}}$) & 4 \\
    Dimension of attention layers ($d^{\text{model}}$) & 256 \\
    Dimension of feed-forward layers ($d^{\text{ff}}$) & 2048 \\
    Number of self-attention blocks ($N$) & 6 \\
    Batch size & 24\\
    Learning rate & 5e-5\\
    Optimization technique & Adam\\
    \bottomrule
    \toprule
    Configuration of mask estimation network\\
    \midrule
    Number of channels in bottleneck (B) & 256\\
    Number of channels in conv blocks (H) & 512\\
    Number of conv blocks in each repeat (X) & 8\\
    Number of repeats (R) & 4\\
    Batch size & 24\\
    Learning rate & 1e-4\\
    Optimization technique & Adam\\
    \bottomrule
    \toprule
    Configuration of STFT\\
    \midrule
    Sampling frequency & 16 kHz\\
    Frame length & 64 ms\\
    Frame shift & 16 ms\\
    Window function & Hanning\\
    \bottomrule
  \end{tabular}
  }
\vspace{-2mm}
\end{table}

\begin{table*}[t]
  \renewcommand{\arraystretch}{1.0}
  \caption{SDR [dB], PESQ, STOI (higher is better), and WER [\%] (lower is better) for non-moving and moving source datasets}
  \label{tab:result}
  \centering
  \scalebox{1.1}{
  \begin{tabular}{ l | c c c c | c c c c }
    \toprule
    & \multicolumn{4}{c |}{non-moving source} & \multicolumn{4}{c}{moving source} \\
    Method & SDR $\uparrow$ & PESQ $\uparrow$ & STOI $\uparrow$ & WER $\downarrow$ & SDR $\uparrow$ & PESQ $\uparrow$ & STOI $\uparrow$ & WER $\downarrow$ \\
    \midrule
    mixture & 5.3 & 1.37 & 0.87 & 4.9 & 5.3 & 1.38 & 0.87 & 4.9\\
    \midrule
    masking & 14.8 & 2.40 & 0.95 & 5.7 & 14.7 & 2.40 & 0.95 & 5.8\\
    tiv\_mvdr & 15.1 & 2.31 & 0.96 & \bf{2.9} & 11.4 & 2.14 & 0.93 & 3.8\\
    onl\_mvdr & 13.5 & 2.24 & 0.95 & 3.4 & 10.2 & 2.08 & 0.92 & 4.1\\
    blk\_mvdr & 13.0 & 2.19 & 0.95 & 3.1 & 11.4 & 2.11 & 0.93 & 3.8\\
    \midrule
    Proposed att\_mvdr & \bf{17.8} & \bf{2.73} & \bf{0.97} & 3.4 & \bf{16.7} & \bf{2.69} & \bf{0.96} & 3.8\\
    + weight\_smooth & 15.4 & 2.48 & 0.96 & 3.0 & 13.9 & 2.48 & 0.95 & \bf{3.4}\\
    \bottomrule
  \end{tabular}
  }
\vspace{-2mm}
\end{table*}

\subsection{Experimental results for moving and non-moving source datasets}

Here, we compare our proposed self-attention-based time-varying MVDR beamformer (att\_mvdr) with time-invariant (tiv\_mvdr), online (onl\_mvdr), and blockwise (blk\_mvdr) MVDR beamformers on moving and non-moving source datasets.
As a comparison, we also provide the results obtained by applying the time-frequency mask to the mixture without any beamforming (i.e., masking).
In this experiment, all of the above mask-based beamformers are constructed with the same estimated time-frequency masks, which are estimated by the time-frequency mask estimation network in Section~\ref{sec:expconf}.
To tune the forgetting factor $\alpha$ and block size parameter $L$ for the online and blockwise MVDR implementations, we 
preliminarily evaluated the enhancement performance for $\alpha = \{0.999, 0.99, 0.9, 0.7, 0.5\}$ and $L = \{5, 10, 20, 30, 40, 50\}$, respectively.
We set the forgetting factor $\alpha$ in Eq~\eqref{eq:onl} to 0.999 and the block size parameter $L$ in Eq~\eqref{eq:blk} to 50 for the moving source dataset, and we set the forgetting factor $\alpha$ to 0.999 and the block size parameter $L$ to 50 for the non-moving source dataset, as they achieved the best WER scores on the development set.
Moreover, we set the number of frames for weight smoothing $L^{'}$ in Eq~\eqref{eq:smooth} to 7 for the moving source dataset and to 9 for the non-moving source dataset.
Table~\ref{tab:result} shows the speech enhancement (i.e., SDR, PESQ, STOI) and ASR (i.e., WER) performance measures for the non-moving and moving source datasets.

First, the left side of Table~\ref{tab:result} shows the results for the non-moving source dataset.
We observe that masking and all conventional variants of MVDR improve the speech enhancement measures, i.e., SDR, PESQ, and STOI.
For ASR, masking degrades performance, probably because it induces distortions that are harmful to ASR \cite{chen2018building}.
All conventional beamformers improve ASR, and the best performance is obtained with tiv\_mvdr.
This result is reasonable because the RIRs are static for this dataset. 

The proposed att\_mvdr achieves higher SDR, PESQ, and STOI scores compared to tiv\_mvdr and comparable WER score when applying the smoothing scheme of Eq \eqref{eq:smooth} (i.e., att\_mvdr+smooth).
This result suggests that even for non-moving situations, the proposed method can improve the computation of the SCMs, probably because it may better adapt to changing noise conditions \cite{kubo2019mask}.

In the second experiment, we investigated the behavior of the proposed approach in the moving source scenario.
The results are shown on the right side of Table~\ref{tab:result}.
We observe that the performance of tiv\_mvdr degrades significantly compared to the non-moving case, i.e., SDR degrades by 3.7 dB and there is a relative WER degradation of more than 20 \%. This result confirms the importance of considering source movements in the design of a beamformer.
onl\_mvdr and blk\_mvdr achieve time-varying beamforming, but they do not contribute to improving speech enhancement and ASR scores compared to tiv\_mvdr. This illustrates the difficulty of setting appropriate hyperparameters to effectively track the moving sources.
In contrast, the proposed att\_mvdr successfully achieved higher SDR, PESQ, and STOI scores compared to tiv\_mvdr.
In addition, by applying the weight smoothing scheme of Eq \eqref{eq:smooth}, the proposed system (i.e., att\_mvdr+weight\_smooth) also successfully improved the WER performance compared to the baseline systems.

These results confirm that the proposed time-varying beamforming approach can mitigate the performance degradation caused by moving sources.

\subsection{Experimental analyses for behavior of proposed self-attention-based time-varying beamformer}

In the following experiments, we analyze the behavior of the proposed scheme.
In these analyses, we used the oracle time-frequency masks to focus on the behavior of the attention weight estimation module.

\subsubsection{Visualization of attention weights}
\label{sec:attweight}

\begin{figure}[t]
\begin{minipage}[b]{0.98\linewidth}
  \centering
  \centerline{\includegraphics[width=0.8\linewidth]{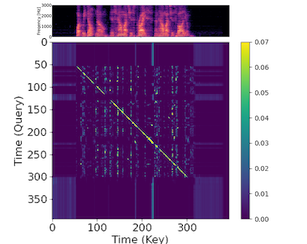}}
  \centerline{(1) Attention weights for speech SCM computation}\medskip
\end{minipage}
\hfill
\begin{minipage}[b]{0.98\linewidth}
  \centering
  \centerline{\includegraphics[width=0.8\linewidth]{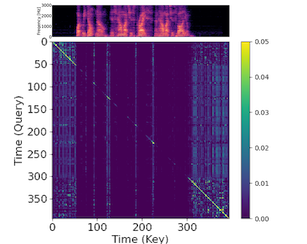}}
  \centerline{(2) Attention weights for noise SCM computation}\medskip
\end{minipage}
%
\caption{Visualization of attention weights in a moving source case}
\label{fig:attweight}
\vspace{-0.3cm}
\end{figure}

We analyzed the behavior of our proposed self-attention-based time-varying attention weight estimation by visualizing the attention weights.
Figure~\ref{fig:attweight} plots the attention weights of an utterance in the moving source dataset for (1) the speech and (2) the noise SCM estimations, respectively.
If the value of the time-frequency masks $m_{t,f}^{\nu}$ is close to zero, the attention weights $c_{t,t^{'}}^{\nu}$ can take arbitrary values without impacting the SCM computation.
This makes it difficult to visualize the attention behavior.
To alleviate this issue, we plot the value of the attention weight multiplied by the voice activity probability $\bar{c}_{t,t^{'}}^{\nu}$ defined as: 
\begin{align}
    \bar{c}_{t,t^{'}}^{\nu} = \frac{c_{t,t^{'}}^{\nu} \bigl( \sum_{f=1}^{F} m_{t^{'},f}^{\nu} \bigr)}{\sum_{\tau=1}^{T} c_{t,\tau}^{\nu} \Bigl( \sum_{f=1}^{F} m_{\tau,f}^{\nu} \Bigr)}.
\end{align}
Figure~\ref{fig:attweight} also plots the spectrogram of the reference clean signal to show the speech activity of the visualized utterance.

We observe from Figure~\ref{fig:attweight}-(1) that for the speech SCM computation the attention weights mainly take high values for the diagonal region at the speech-active time steps. This means that the speech SCM computation focuses on the ISCMs around the time index of the query, i.e., $t$ in $c^{\nu}_{t,t^{'}}$. 
In contrast, for the noise SCM computation, the attention module mainly focuses on the speech-inactive regions (e.g., beginning and end parts of the utterance), regardless of the time index of the key as shown in Figure~\ref{fig:attweight}-(2). 

This behavior seems reasonable because in the dataset the speaker moves while the noise signal consists of non-moving diffuse noise.

\subsubsection{Visualization of beam patterns}
\label{sec:beampattern}

\begin{figure}[t]
  \centering
  \includegraphics[width=0.6\linewidth]{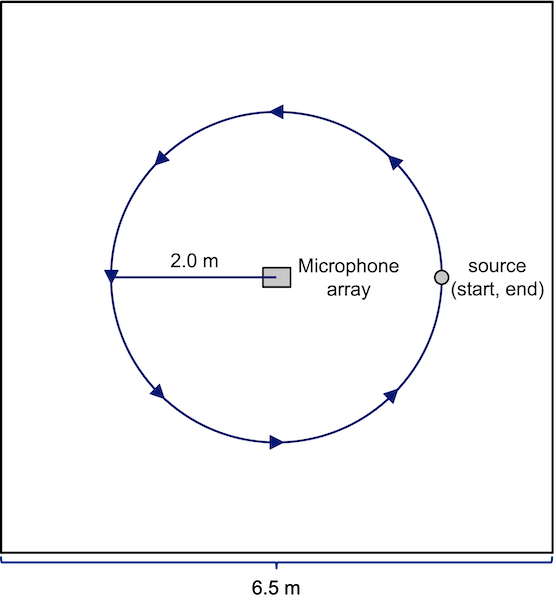}
  \caption{Simulation layout of microphone array and moving source (moving in circles) for the analysis of the beam patterns of Section~\ref{sec:beampattern}}
  \label{fig:simulation2}
\end{figure}

\begin{figure*}[t]
  \centering
  \includegraphics[width=1.0\linewidth]{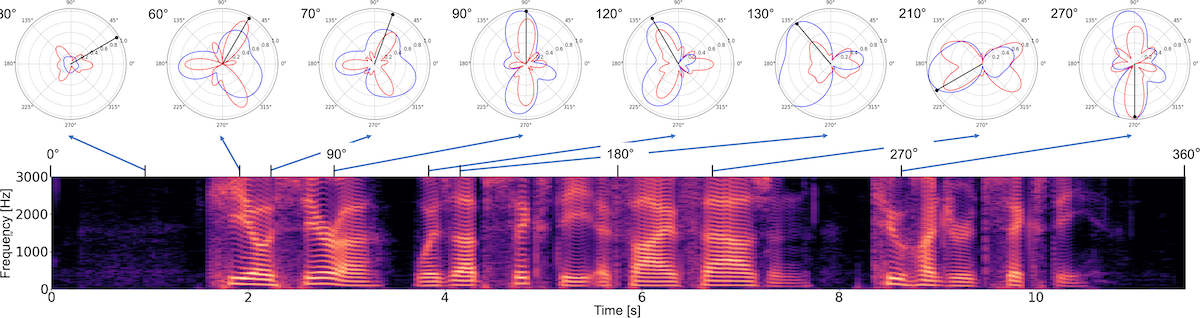}
  \caption{Visualization of beam patterns for a source moving around a circle as shown in Figure~\ref{fig:simulation2}}
  \label{fig:beampattern}
\end{figure*}

Next, we analyze the behavior of the time-varying beamforming filters estimated by the proposed method by visualizing the beam patterns of the constructed MVDR beamformer.
To emphasize the source movement, we consider here a source moving on a circle in the room, as illustrated in Figure~\ref{fig:simulation2}.
The evaluated utterance is thus simulated under a different RIR configuration from the training utterances in the moving source dataset.
The room width and depth are set to 6.5 m, and the room height is set to 3.0 m.
The number of trajectory points is set to 360.

Figure~\ref{fig:beampattern} shows an example of the beam patterns for a moving source, where beam patterns at eight time frames are shown.
The black straight line denotes the actual direction of the source speaker at that frame, i.e., 30$^\circ$, 60$^\circ$, 70$^\circ$, 90$^\circ$, 120$^\circ$, 130$^\circ$, 210$^\circ$, and 270$^\circ$.
The blue and red lines correspond to the beam patterns for 1 kHz and 2 kHz, respectively.

The first beam pattern on the left (i.e., 30$^\circ$) corresponds to a region where the source is inactive. In this case, the beamformer does not show any clear directivity pattern.
On the other hand, when the source speaker is active (e.g., 60$^\circ$ and 90$^\circ$), we can confirm that the beamformer has a main lobe toward the direction of the source speaker.
Moreover, we observe that the beam patterns change over time and follow the source positions.
These visualizations suggest that the estimated beamforming filters of the proposed method can successfully track the positions of a moving source speaker.

\section{CONCLUSION}
\label{sec:conclusion}

In this paper, we discussed the application of mask-based beamformers to moving source situations. We introduced a generalized view of conventional approaches for computing the SCMs of moving sources, which can be interpreted as a sum of ISCMs weighted by attention weights. We proposed using an NN to compute these attention weights and showed that the self-attention-based NN is a reasonable candidate for this task.

We performed experiments showing the impact of moving sources on conventional beamformers.
The results show that it was challenging to achieve high enhancement and ASR performance when a source was moving even with an online or blockwise implementation of the mask-based beamformer.
In contrast, the proposed scheme uses an NN to predict optimal attention weights to compute the time-varying SCMs.
This resulted in stable performance for both moving and non-moving conditions.

These results demonstrate the potential of our proposed approach as well as the importance of addressing the moving source conditions.
Future works should include application of this framework to more challenging conditions such as dealing with moving interfering sources, as well as extend the approach to low-latency processing by, for example, reducing the scope of the attention computation to the past samples.


\bibliographystyle{IEEEtran}
\bibliography{refs}

\end{document}